\documentclass[proof]{WileyASNA-v1}

\articletype{ORIGINAL ARTICLE}%

\received{xx xxxx xx}
\revised{xx xxxx xx}
\accepted{xx xxxx xx}

\begin{document}

\title{Measuring the precise photometric period of the probable intermediate polar
1RXS J014549.6+514314 based on extensive photometry}
%\protect\thanks{This is an example for title footnote.}}

\author{V. P. Kozhevnikov}

\authormark{KOZHEVNIKOV\textsc{}}

\address{\orgdiv{Astronomical Observatory}, \orgname{Ural Federal University}, \orgaddress{\state{Yekaterinburg}, \country{Russia}}}

\corres{V. P. Kozhevnikov, Ural Federal University, 19 Mira Street, 620002  Yekaterinburg, Russia \email{valery.kozhevnikov@urfu.ru}}

%\presentaddress{This is sample for present address text this is sample for present address text}

\abstract{ 
Recently, in the VSX database, G.~Murawski reported the discovery of an oscillation with a period of about 40~min in the probable cataclysmic variable 1RXS~J014549.6+514314. To confirm the existence of this oscillation and precisely measure its period, I conducted extensive photometric observations of the object over 25 nights between 2022 and 2024. The total duration of these observations was 139~hr. The observations revealed the oscillation on each observation night. This oscillation was coherent throughout all the observations covering 18 months. Due to the large coverage of observations, I was able to determine the oscillation period with high precision. It was found to be $41.485867\pm0.000026$~min. The semi-amplitude of the oscillation was large, averaging 0.154(2)~mag. The pulse profile of the oscillation did not change during the observations and followed a perfect sinusoidal pattern. With the high precision of the period, I created an ephemeris for the oscillation with a validity time of 30 years. This ephemeris can be used in future studies of changes in the oscillation period. Although short-period X-ray oscillations have not yet been detected in 1RXS J014549.6+514314, the high coherence of its 41-minute oscillation suggests that this object is most likely to be an intermediate polar.}

\keywords{Stars: individual, 1RXS J014549.6+514314; Novae, Cataclysmic variables; Stars: oscillations}

\jnlcitation{\cname{%
\author{Kozhevnikov V. P.}, 
} (\cyear{2025}), 
\ctitle{Measuring the precise photometric period of the probable intermediate polar 1RXS J014549.6+514314 based on extensive photometry}, \cjournal{Astron. Nachr.}, \cvol{xxxx;xx:x-x}.}

%%\fundingInfo{Funding info text.}

\maketitle

%\footnotetext{\textbf{Abbreviations:} ANA, anti-nuclear antibodies; APC, %antigen-presenting cells; IRF, interferon regulatory factor}

\section{Introduction}\label{sec1}

Intermediate polars (IPs) are a subclass of cataclysmic variables (CVs) and represent a group of binary systems in which a magnetic white dwarf accretes matter from a late-type stellar companion, which fills its Roche lobe. In these systems, the spin of the white dwarf is asynchronous with the orbital motion, leading to accretion through a truncated accretion disc in most cases. These characteristics give rise to a multitude of phenomena associated with periodic variability. Due to the tilt of the magnetic dipole axis of the white dwarf with respect to its spin axis, a primary oscillation occurs with the spin period of the white dwarf. This oscillation is observed in both optical and X-ray emissions.The reprocessing of X-rays by the secondary star or asymmetric regions of the accretion disc generates an optical oscillation with a beat frequency defined as $1/P_{\rm beat} = \omega-\Omega$, where $\omega=1/P_{\rm spin}$ and  $\Omega=1/P_{\rm orb}$. This represents the lower orbital sideband of the spin frequency. Although the secondary star and the accretion disc are unable to effectively reflect X-rays with this beat frequency as described by \citet{norton92}), nonetheless, a robust X-ray oscillation with a frequency of $\omega - \Omega$ can arise due to disc overflow accretion, as reported by \citet{hellier91}, \citet{norton92} and \citet{hellier14}. Furthermore, oscillations with frequencies $\omega-2\Omega$, $\omega + \Omega$, and $\omega+2\Omega$ may be observed due to amplitude modulation caused by geometrical effects. However, these signals are not as strong as the strength of $\omega$ or $\omega - \Omega$, as noted by \citet{warner86}. Reviews of IPs can be found in \citet{patterson94, warner95, hellier96, hellier01} and \citet{hellier14}. The properties of X-rays from IPs are described in \citet{kuulkers06} and \citet{mukai17}. The most recent information about IPs can be accessed through \citet{demartino20}.

Coherent X-ray oscillations with periods significantly shorter than the orbital period are a defining characteristic of IPs. 
This is strong evidence for classifying a star as an IP \citep{mukai05}. However, optical short-period oscillations with high stability and the presence of one or more orbital sidebands also support inclusion in the list of IPs, even if there are no X-ray detections \citep{kuulkers06}. A simple and effective way to differentiate this stable oscillation from intermittent, quasi-periodic variations is by checking whether the same period can be detected in multiple data sets \citep{mukai17}. This also means that the oscillation can be observed with observations of comparable quality.

At large time scales, the spin periods of IPs can experience small variations due to interactions between braking and accretion torques \citep[e.g.,][]{warner91}. Long-term monitoring of spin periods allows us to verify spin equilibrium by comparing observed and calculated maximum spin pulse times using a precise oscillation ephemeris, or by directly measuring the white dwarf spin period \citep{patterson20}. Spin equilibrium can be verified by observing alternating periods of spin-up and spin-down, which occur as a result of changes in the accretion rate \citep{patterson94}. This spin equilibrium check is significant because many theoretical studies assume that IPs are in a state of equilibrium \citep[e.g.,][]{norton04}. Observations of continuous spin-up or spin-down also play a crucial role, as they provide insight into the angular momentum transfer within the binary system \citep{king99}.

The source 1RXS J014549.6+514314 (hereafter J0145), which was listed in the ROSAT All-Sky Bright Source Catalogue by \citet{voges99}, was identified by \citet{haakonsen09} as a point-like object with a J-band magnitude of 16.7, and given the designation 2MASS J01454839+5143187. Based on the ASAS-SN Sky Patrol Light Curve {\small (\url{https://asas-sn.osu.edu/sky-patrol/coordinate/e578ba6f-adeb-4790-b748-828b4bf605a2})} T. Kato concluded that this object is likely a CV {\small (\url{http://ooruri.kusastro.kyoto-u.ac.jp/mailarchive/vsnet-chat/8145})}. A periodic brightness oscillation with a period of about 40~min was discovered in J0145 by G.~Murawski using observations from the ZTF photometric survey (\url{https://irsa.ipac.caltech.edu/Missions/ztf.html}) and an observer with the code "NADA". This was reported in the International Variable Star Index (VSX) database (\url{https://www.aavso.org/vsx/}). E.~Breedt obtained the optical spectrum of this object, which revealed distinct hydrogen and helium emission lines {\small (\url{https://vsx.aavso.org/vsx_docs/844443/4999/1RXS0145%2B5143.pdf})}. Based on this information, we can conclude that J0145 is likely a CV and the detected period may be related to the spin period of the white dwarf. Therefore, J0145 could probably be classified as an IP.

Unfortunately, in the VSX database, G.~Murawski did not provide a detailed description of the data or analysis that led to the detection of this oscillation with a period of about 40~min, nor did he provide an estimate for the period error. To confirm the existence of this oscillation in J0145, I conducted photometric observations using a 70-cm telescope at the Kourovka Observatory of the Ural Federal University. The oscillation was easily detected in the light curves obtained from these observations. To ascertain the coherence of this oscillation and determine its precise period, as well as to obtain a long-term ephemeris for the oscillation, extended observations were conducted over 25 nights covering a total of 139 hours over 18 months. This paper presents the results of these observations.

%Voges, W. ; Aschenbach, B. ; Boller, Th. ; Br?uninger, H. ; Briel, U. ; Burkert, W. et al. Astronomy and Astrophysics, v.349, p.389-405, 1999
%Christian Bernt Haakonsen and Robert E. Rutledge The Astrophysical Journal Supplement Series, 184:138–151, 2009 
%\url{https://asas-sn.osu.edu/sky-patrol/coordinate/e578ba6f-adeb-4790-b748-828b4bf605a2}
%\url{http://ooruri.kusastro.kyoto-u.ac.jp/mailarchive/vsnet-chat/8145}
%\url{https://www.aavso.org/vsx/}
%2MASS J01454839+5143187     из каталога 01454839+5143187
%https://vsx.aavso.org/vsx_docs/844443/4999/1RXS0145%2B5143.pdf (1RXS0145+5143)
%https://irsa.ipac.caltech.edu/Missions/ztf.html

\section{Observations}\label{sec2}

For observations of variable stars, I use a three-channel photometer equipped with photomultiplier tubes. This device converts photons into electrical pulses, allowing for accurate counting of photons and measurement of light intensity. The design and noise analysis of the photometer can be found in \citet{kozhevnikoviz00}. It continuously monitors the brightness of two stars and the sky background within the field of view of a telescope, providing high precision even under challenging conditions such as unstable atmospheric transparency and varying sky backgrounds.

Three pulse counters transmit data to a computer, which processes information using specialized software. The photometer works in conjunction with a 70-cm Cassegrain telescope equipped with computer-controlled stepping motors and a CCD guiding system. This guiding system corrects tracking errors and centers each of the two stars within the photometer's own diaphragm during observations. All components, including the telescope and the photometer, operate automatically under computer control, but human intervention may be required if thick clouds interfere with the observation process.

\begin{table}[t]
\caption{Log of the observations.}
\label{tab1}
\begin{tabular}{@{}l c c c }
\hline
\noalign{\smallskip}
Date (UT) & BJD$_{\rm TDB}$ start & Length    & Magnitude   \\ 
                 & (-2400000)                   &   (h)        &      \\
\noalign{\smallskip}
\hline
\noalign{\smallskip}
2022 Aug. 23    &   59815.281534    &   2.1   &     15.916(10)  \\
2022 Aug. 24    &   59816.234009    &   5.1   &     16.228(5)  \\
2022 Aug. 25    &   59817.229931    &   5.4  &      16.062(5)  \\
2022 Aug. 26    &   59818.266491    &   4.6   &     16.209(6)  \\
2022 Aug. 27    &   59819.229113    &   2.1   &     16.154(10)  \\
2022 Aug. 28    &   59820.228111    &   5.7  &      16.054(6)  \\
2022 Aug. 29    &   59821.216836    &   5.9   &     16.123(4)  \\
2022 Nov. 27    &   59911.202917    &   9.0  &      16.037(4)  \\
2022 Nov. 28    &   59912.098374    &   9.0  &      15.950(4)  \\
2022 Dec. 2      &   59916.411256    &   4.4  &      15.953(4)  \\
2022 Dec. 21    &   59935.080264    &   6.3  &      15.966(4)  \\
2022 Dec. 28    &   59942.065537    &   3.6   &     15.903(8)  \\
2022 Dec. 29    &   59943.064302    &   9.4   &     15.829(6)  \\
2023 Jan. 13    &   59958.109414    &   4.4  &      15.841(6)  \\
2023 Jan. 15    &   59960.156868    &   6.4   &     15.825(5)  \\
2023 Jan. 16    &   59961.078889    &   7.1   &     15.923(5)  \\
2023 Jan. 17    &   59962.152034    &   6.5  &      15.873(5)  \\
2023 Jan. 18    &   59963.112192    &   5.7   &     15.876(6)  \\
2023 Jan. 19    &   59964.105479    &   8.8  &      15.987(4)  \\
2023 Oct. 16    &   60234.328539    &   1.8  &      16.045(6)  \\
2023 Nov. 6     &   60255.087533    &   5.9  &      16.230(6)   \\
2024 Jan. 6     &   60316.073657    &   3.0  &      16.283(7)   \\
2024 Feb. 14   &   60355.118448    &   6.4  &      16.243(6)   \\
2024 Feb. 15   &   60356.121800    &   5.9  &      16.123(7)   \\
2024 Feb. 18   &   60359.126875    &   4.2  &      16.278(9)   \\
\noalign{\smallskip}
\hline
\end{tabular} 
\end{table}

I performed photometric observations of J0145 using white light without filters (approximately 3000-8000\AA) and a time resolution of 16~s. Although this time resolution may seem excessively high for analyzing the oscillation reported by G.~Murawski, which has a period of about 40~min, it allows me to accurately fill in gaps between individual observation nights. For each of the two stars, I measured the light intensity using a 16-arcsec diaphragm. To reduce photon noise caused by the sky background, I used a 40-arcsec diaphragm for the sky. 

The comparison star was $Gaia$ DR3 740057406155138560 with $G = 11.84$~mag and $BP - RP = 0.96$~mag, and its color index was similar to that of J0145 ($Gaia$ DR3 406158049112734336) with $G = 16.33$~mag and $BP - RP = 0.47$~mag \citep{gaia16, gaia23}. The similarity in the color indexes of these two stars reduces the effect of differential extinction. Differential magnitudes were calculated by taking into account the differences in photomultiplier tube sensitivities and the differences in sky background measurements caused by different diaphragm sizes.

%1RXS J0145  DR3 406158049112734336   G=16.33 0.47
%CP4  11.84  0.96  DR3 740057406155138560   0.96-0.47=0.49

On the first observation night, I found that J0145 was quite bright, although it was invisible to the eye (about 16~mag). Because the total light intensity of J0145 and the sky background were noticeably higher than the sky background alone, it was easy to find the star within the photometer diaphragm by adjusting the telescope position according to stellar coordinates and observing the photometer counts. Then the centring of the star was improved by means of small displacements of the telescope along right ascension and declination. That night, in J0145, I detected an oscillation with a period of about 40~min, which was reported by G.~Murawski. This oscillation was clearly visible in the light curve. To measure its period with high precision, I conducted prolonged observations spanning 139 hours over 25 nights. The observations covered a time span of 18 months. The differential light curves with a time resolution of 16~s are available at the following link for further analysis: \url{https://www.researchgate.net/publication/396463835}. The observational log is given in Table~\ref{tab1}.

%2.1+5.1+5.4+4.6+2.1+5.7+5.9+9.0+9.0+4.4+6.3+3.6+9.4+4.4+6.4+7.1+6.5+5.7+8.8+1.8+5.9+3.0+6.4+5.9+4.2=138.7
%Mean magnitude 16.04pm0.03  st. dev.=0.15 changes 15.9-16.3 SD=0.1
%Mean period 41.27pm0.10  st. dev.=0.5 
%Mean semiamplitude 0.164pm0.006  st. dev.=0.029

\begin{figure}[t]
\centering
\includegraphics[width=84mm]{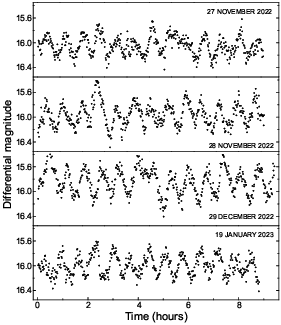}
\caption{Longest light curves of J0145, in which the oscillation with a period of about 40~min is clearly visible.} 
\label{fig1}
\end{figure}

\section{Analysis and results}\label{sec3}

The photon counts should follow a Poisson distribution with $N=\sigma^2$. Here, $N$ is the average number of photons and $\sigma^2$ is the variance of the data points in the time series. I verified this property using a stable light source for my photomultiplier tubes, finding that $N/\sigma^2$=0.98$\pm$0.01. This allows me to estimate the photon noise in the light curves of J0145. To calculate the photon noise for one of the star channels, I used a formula provided by \citet{kozhevnikov25}. During observations of J0145, the photon noise of differential light curves obtained with a time resolution of 16~s ranged from 0.06 to 0.08~mag.
%foton 16 s 0.062 = 0.079 mag          
%foton x 4 (64s) 4^0.5=2 0.031 - 0.040m

In Fig.~\ref{fig1} and other relevant locations, I converted differential magnitudes to $G$-band magnitudes using the $G$-band magnitude of the comparison star. The $G$-band used in the Gaia mission closely matches the wavelength range that I use in my observations with photomultiplier tubes that have S20 photocathodes. This similarity makes my white-light magnitude scale similar to the $G$-band magnitude scale.

%1RXS J0145  DR3 406158049112734336   G=16.33 0.47
%CP4  11.84  0.96  DR3 740057406155138560   0.96-0.47=0.49

To better display the observed oscillation, Fig~\ref{fig1} presents the longest light curves of J0145 with a time resolution of 64~s. The photon noise in these light curves is reduced by a factor of 2 compared to that obtained with a time resolution of 16~s. As seen, these light curves clearly exhibit the oscillation with a period of about 40~min. In addition, it can be seen that the average brightness of J0145 in these light curves is approximately the same. To determine the overall brightness behaviour, I averaged the magnitudes of J0145 for each light curve obtained and included them in Table~\ref{tab1}. Based on these measurements, it appears that during the 18 months of observations, the brightness of J0145 varied between 15.8--16.3~mag and there were no significant changes resembling dwarf nova outbursts.

My multi-channel photometer allows me to obtain measurements with precisely equal time intervals between them. Data collected at regular time intervals can best be analyzed using methods based on the fast Fourier transform (FFT). Indeed, FFT enables extremely rapid calculations, allowing the analysis of combined time series with high temporal resolution, covering time spans of up to a year or more. Additionally, due to its fundamental functions, namely, sine and cosine, Fourier analysis is well-suited to signals that exhibit short-period oscillations with smooth, quasi-sinusoidal waveforms, such as in the case of an IP.

Most FFT algorithms require that the time series contains a number of data points that is a power of two, $N=2^p$, where $p$ is an integer. The number of frequency components in the resulting power spectrum is then $N/2$. In this case, maximum frequency resolution is achieved, $\Delta f=1/T$, where $T$ is the length of the time series \citep[e.g.,][]{bendat86}. This power spectrum covers all frequencies between the lowest frequency $1/T$ and the Nyquist limit $1/2t$, where $t$ is time resolution. The frequency components of a classical Fourier power spectrum are calculated by taking the sum of the squared real and imaginary parts of a Fourier transform multiplied by $2/N$, where $N$ is the number of data points in the time series \citep[e.g.,][]{bendat86}. However, when calculating the power spectrum, I multiply this sum by $2/N^2$, resulting in a power spectrum with a simple interpretation. In this power spectrum, the height of the peak caused by a coherent oscillation is equal to $A^2/2$, where $A$ is the semi-amplitude of this oscillation. The sum of all frequency components is equal to the variance of the time series, as follows from Parseval's theorem \citep[see, e.g.,][]{kozhevnikoviz00}. The latter implies that as the number of data points in a time series increases, the level of noise in its power spectrum decreases.

\begin{table}[t]
\caption{The oscillation periods and semi-amplitudes obtained from different nights.}
\label{tab2}
\begin{tabular}{@{}l c c c }
\hline
\noalign{\smallskip}
Date (UT) & Length    &  Period  & Semi-amplitude \\ 
                 &   (h)        &     (min  ) &   (mag)    \\
\noalign{\smallskip}
\hline
\noalign{\smallskip}
2022 Aug. 23    &   2.1   &     41.6 &  0.16 \\
2022 Aug. 24    &   5.1   &     40.8 &  0.11 \\
2022 Aug. 25    &   5.4  &      41.6 &  0.17 \\
2022 Aug. 26    &   4.6   &     41.2 &  0.16 \\
2022 Aug. 27    &   2.1   &     40.1 &  0.18  \\
2022 Aug. 28    &   5.7  &      41.2 &  0.21 \\
2022 Aug. 29    &   5.9   &     42.0 &  0.12 \\
2022 Nov. 27    &   9.0  &      41.6 &  0.14 \\
2022 Nov. 28    &   9.0  &      41.2 &  0.18 \\
2022 Dec. 2      &   4.4  &      41.2 &  0.16 \\
2022 Dec. 21    &   6.3  &      40.8 &  0.16 \\
2022 Dec. 28    &   3.6   &     41.2 &  0.20 \\
2022 Dec. 29    &   9.4   &     41.6 &  0.21 \\
2023 Jan. 13    &   4.4  &      40.8 &  0.20 \\
2023 Jan. 15    &   6.4   &     41.6 &  0.15 \\
2023 Jan. 16    &   7.1   &     41.6 &  0.16 \\
2023 Jan. 17    &   6.5  &      41.2 &  0.17 \\
2023 Jan. 18    &   5.7   &     41.2 &  0.20 \\
2023 Jan. 19    &   8.8  &      41.6 &  0.17 \\
2023 Oct. 16    &   1.8  &      41.6 &  0.12 \\
2023 Nov. 6     &   5.9  &      41.2 &  0.21 \\
2024 Jan. 6     &   3.0  &      41.2 &  0.14 \\
2024 Feb. 14   &   6.4  &      40.1 &  0.14 \\
2024 Feb. 15   &   5.9  &      41.2 &  0.16 \\
2024 Feb. 18   &   4.2  &      42.4 &  0.13 \\
\noalign{\smallskip}
\hline
\end{tabular} 
\end{table}

%average period all 41.27pm0.10  average amplitude all = 164pm6 mmag   
%average amplitude 16 = 168pm8 srsp16: 0.1181*1.414=0.167

The precision of determining the oscillation frequency is not equal to the frequency resolution, and it can be much higher. It depends on both the frequency resolution and the noise level in the power spectrum. \citet{schwarzenberg91} demonstrated that the $1\sigma$ confidence interval for the oscillation frequency is equal to the width of the peak in the power spectrum at the $S-N$ level. Here, $S$ is the height of the peak and $N$ is the average level of noise around the peak. To precisely find the maximum and width of the peak, the number of frequency components must be significantly increased. This can be achieved by adding zeros to the end of a time series. This decreases the frequency step, but does not change the frequency resolution. The new frequency step is then $\Delta f=1/T_{\rm total}$, where $T_{\rm total}$ is the original length of the time series plus the length corresponding to the added zeros. For a combined time series, the signal-to-noise ratio of the peak in the power spectrum can exceed 100. In this case, the number of frequency components in the entire power spectrum can be very large, reaching $2^{25}$. And the peak itself may contain more than 100 frequency components. However, with FFT, an ordinary computer can perform these calculations in a minute.

As shown in Figure~\ref{fig1}, there is a clear oscillation with a period of about 40~min in the individual light curves. Therefore, I determined its period and semi-amplitude for each of the 25 observed light curves using the power spectra. Before this, variations with periods longer than the length of the light curve were subtracted by fitting a first- or second-order polynomial, which is a common procedure in Fourier analysis \citep[e.g.,][]{bendat86}. The results of these measurements are presented in Table~\ref{tab2}. No trends in the measured periods or amplitudes were observed over time. The average period was found to be 41.27(10)~min, and the average semi-amplitude was 0.164(6)~mag, with errors determined as the root mean square (RMS) errors of the averages.

\begin{figure}[t]
\centering
\includegraphics[width=84mm]{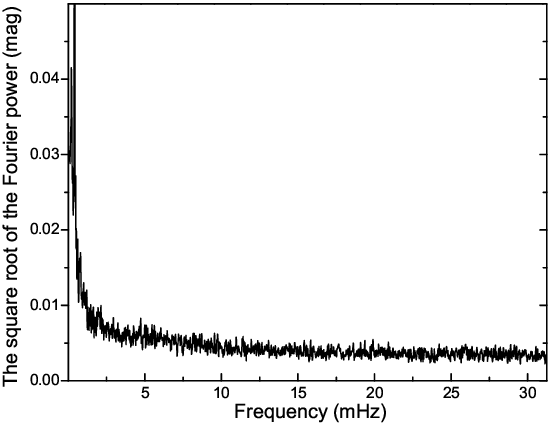}
\caption{Averaged power spectrum calculated from 16 individual power spectra of nights that were longer than 5 hours each. In addition to the large peak corresponding to a period of 41~min seen on the left, truncated at the 30 per cent level, there is red noise at frequencies up to 20~mHz.}
\label{fig2}
\end{figure}

As seen in Figure~\ref{fig1}, some of the cycles of the 41-min oscillation are significantly distorted. These distortions may be caused by the flickering of J0145. To detect and characterize this flickering, I calculated the averaged power spectrum using 16 light curves longer than 5~hr. This spectrum is shown in Figure~\ref{fig2}. In addition to the peak corresponding to the 41-min oscillation, there is red noise at frequencies reaching at least 20~mHz, which indicates the flickering of J0145. While red noise can also be caused by atmospheric effects, atmospheric red noise frequencies in my three-channel photometer do not exceed 2~mHz and their semi-amplitudes at low frequencies do not exceed a few thousandths of a magnitude \citep[see Figure 2 in][]{kozhevnikoviz00}. So, as follows from Figure~\ref{fig2}, the flickering of J0145 is strong, reaching high frequencies up to 20~mHz with large semi-amplitudes of tenths of a magnitude at low frequencies.

Using the averaged power spectrum shown in Figure~\ref{fig2}, I calculated the oscillation period and its error. The error was calculated using the method proposed by \citet{schwarzenberg91}. Although this method was originally designed for individual power spectra, it can also be applied to averaged power spectra by simply dividing the error by the square root of the number of power spectra used for averaging. To determine the precise maximum and width of the peak at the $S-N$ level, I used a Gaussian fit to the upper part of the peak profile. The profile was calculated using a sufficient number of frequency components. The oscillation period was found to be 41.41(13)~min, which is consistent with the average period determined from individual periods in separate power spectra, which was 41.27(10)~min. Although the error is slightly larger than that obtained from averaging individual periods, this is not surprising given that the length of the longest light curves used is only about 80 per cent of the total length. Moreover, the small difference in these errors confirms that the error calculated by the method proposed by \citeauthor{schwarzenberg91} corresponds to the RMS error.

The appearance of light curves and the presence of repeated peaks in individual power spectra leave no doubt about the reality of the 41-min oscillation. However, the precision of the period determined using individual power spectra is rather low due to the frequency resolution of these power spectra. A much greater frequency resolution can be achieved by combining time series consisting of individual light curves and gaps between them corresponding to the absence of observations. Filling in these gaps with zeros is a reasonable approach, as it avoids discontinuities in the data. Indeed, after removing periods that are longer than the length of individual light curves, the average value for each light curve would be zero.

Due to gaps in the data, in addition to the main peak corresponding to a periodic oscillation, other components may appear in power spectra that are consistent with the window function. However, this can be seen as an advantage rather than a disadvantage. Indeed, these components arise from the coherent phase of an oscillation, and the close similarity between the power spectrum and the window function derived from an artificial time series consisting of a sine wave and gaps, such as those found in observations, proves the coherence of the oscillation.

\begin{figure}[t]
\centering
\includegraphics[width=84mm]{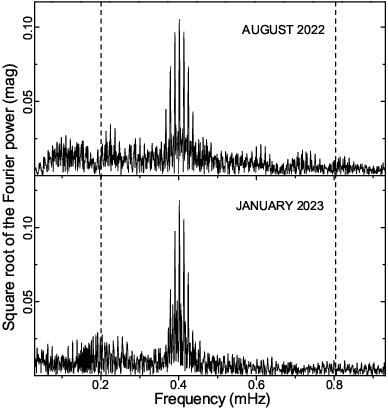}
\caption{Power spectra of two combined time series, consisting mostly of consecutive nights obtained in August 2022 and January 2023. The dashed lines indicate the expected frequencies corresponding to the first subharmonic and the first harmonic of the 41-min oscillation.}
\label{fig3}
\end{figure}

First, I constructed two combined time series, consisting mostly of consecutive nights obtained in August 2022 and January 2013.  The power spectra of these time series have simple window functions. In the case of a coherent oscillation, they should show a main peak and its one-day aliases. Based on observational coverage, the frequency resolution of these time series is about 16 times higher than that of the longest light curve (see Table~\ref{tab1}). To remove the unnecessary high-frequency components and thus reduce the total number of frequency components in the entire power spectrum, I averaged the combined time series over 128-s time intervals. The low-frequency parts of their power spectra are shown in Figure~\ref{fig3}. They reveal distinct main peaks of the 41-min oscillation surrounded by its one-daily aliases. To determine the peak maxima and their widths at $S-N$ levels, I used Gaussian fits to the upper parts of the peak profiles. The profiles were calculated using sufficiently small frequency steps. The oscillation periods were found to be 41.488(12) and 41.486(9)~min, respectively. Although these periods are more precise than the period obtained from the averaged power spectrum, the improvement in precision due to increased frequency resolution is less than expected. This is likely due to the influence of noise, as only 22 and 28 per cent of the total data was used in the first and second cases, respectively.

In Figure~\ref{fig3}, the dashed lines show the frequencies of the first subharmonic and the first harmonic of the 41-min oscillation. There are no noticeable peaks at these frequencies. The absence of subharmonics indicates that the 41-min oscillation is fundamental, not a harmonic of a longer period oscillation. The absence of the first harmonic indicates a simple profile for the pulse of the 41-min oscillation similar to that of a sine wave.

Both spectra shown in Figure~\ref{fig3} reveal many clusters of peaks, where the peaks are distributed in frequency as one-day aliases. In these two spectra, the peaks do not coincide in frequency and therefore do not indicate coherent oscillations. Pure white noise, such as photon noise, cannot create any organized structures. So, these clusters of peaks are probably created by the flickering of the star, producing weakly coherent (or quasi-periodic) oscillations. The origin of these clusters is also due to insufficient frequency resolution, which is caused by the small observation coverage. Thus, the presence of one-day aliases does not provide conclusive evidence of oscillation coherence.

\begin{figure}[t]
\centering
\includegraphics[width=84mm]{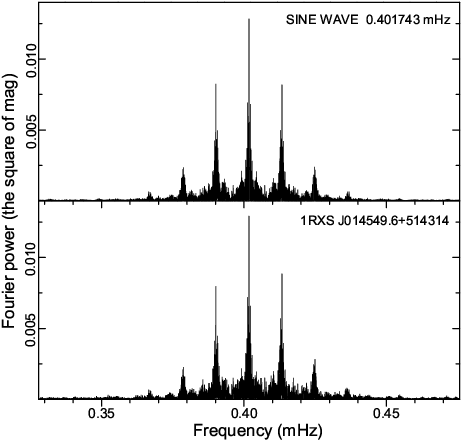}
\caption{Part of the power spectrum of the overall combined time series shown at frequencies around the frequency of the 41-min oscillation (bottom frame). It reveals a structure similar to that of the window function (top frame).}
\label{fig4}
\end{figure}

\begin{figure}[t]
\centering
\includegraphics[width=84mm]{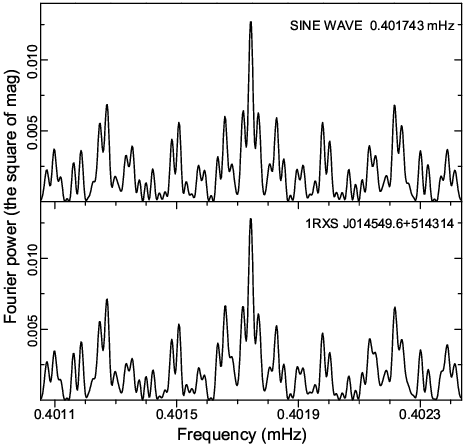}
\caption{Part of the power spectrum of the overall combined time series shown at frequencies around the main peak of the 41-min oscillation on an expanded frequency scale (bottom frame). This power spectrum reveals a fine structure that strikingly coincides with the fine structure of the window function (top frame).}
\label{fig5}
\end{figure}

The overall combined time series has the highest frequency resolution due to its longest observation coverage of 18 months, which is about 1400 times larger than that of the longest light curve (see Table~\ref{tab1}). In addition, the corresponding power spectrum has the lowest level of noise. To remove the unnecessary high-frequency components, I averaged the overall combined time series over 128-s time intervals, as before. Nevertheless, the number of frequency components in the entire power spectrum ensuring sufficiently small frequency steps turned out to be very large and equal to $2^{24}$. Figure~\ref{fig4} shows this power spectrum at frequencies around the frequency of the 41-min oscillation, revealing a complex structure consisting of the main peak and one-day aliases, surrounded by other aliases similar to those seen in the window function on the top frame of Figure~\ref{fig4}. As shown in Figure~\ref{fig5}, when the power spectrum is presented on an expanded frequency scale, its fine structure strikingly coincides with the fine structure of the window function, proving complete coherence of the 41-minute oscillation. Indeed, every detail in this power spectrum arises from the interference between the phases of the 41-min oscillation in each light curve observed.

To determine the peak maximum and its width at the $S-N$ level in the power spectrum of the overall combined time series, I used a Gaussian fit to the upper part of the peak profile, as before. The oscillation period was found to be 41.485849(56)~min. Its precision is about 2300 times better than that obtained using the averaged power spectrum, while the frequency resolution is only about 1400 times higher than that of the longest light curve. Clearly, the increase in precision was achieved not only due to higher frequency resolution, but also due to a lower level of noise in the power spectrum.

%\scriptsize

\begin{table}[t]
\caption[ ]{Verification of the ephemeris.}
\label{tab3}
\begin{tabular}{@{}l c c c }
\noalign{\smallskip}
\hline
\noalign{\smallskip}
BJD$_{\rm TDB}$ max & Semi-ampl.  & Number    &       O--C       \\
    (-2400000)                & (mag)        & of cycles   &        (d)          \\
\noalign{\smallskip}
\hline
\noalign{\smallskip}
59815.34290(50)     & 0.155(17)     & 0        & 0   \\
59816.35013(70)     & 0.098(15)     & 35       & -0.00111    \\
59817.32999(31)     & 0.171(12)     & 69       & -0.00077  \\
59818.36929(39)     & 0.157(13)     & 105     & 0.00138  \\
59819.28822(49)     & 0.166(17)     & 137     & -0.00159   \\
59820.35617(30)     & 0.209(13)     & 174     & 0.00040   \\
59821.33408(46)     & 0.115(12)     & 208     & -0.00122   \\
59911.39290(30)     & 0.141(9)      & 3334    & -0.00126   \\
59912.28606(28)     & 0.176(11)    & 3365    & -0.00120   \\
59916.52203(39)     & 0.157(13)    & 3512    & -0.00025   \\
59935.21939(30)     & 0.150(10)    & 4161    & -0.00032   \\
59942.16084(37)     & 0.207(17)    & 4402    & -0.00199   \\
59943.28638(24)     & 0.216(11)    & 4441    & -0.00003    \\
59958.20893(37)     & 0.176(14)    & 4959    & -0.00086   \\
59960.31331(37)     & 0.155(13)    & 5032    & 0.00042    \\
59961.23585(28)     & 0.157(10)    & 5064    & 0.00106   \\
59962.29986(33)     & 0.159(11)    & 5101    & -0.00090   \\
59963.22436(33)     & 0.195(14)    & 5133    & 0.00169    \\
59964.28797(24)     & 0.167(9)      & 5170    & -0.00065   \\
60234.37688(50)     & 0.113(12)    & 14545   & -0.00190   \\
60255.20781(28)     & 0.210(13)    & 15268   & -0.00032   \\
60316.14152(41)     & 0.142(13)    & 17383   & 0.00106    \\
60355.26186(62)     & 0.114(15)    & 18741   & -0.00207   \\
60356.24316(31)     & 0.174(12)    & 18775   & -0.00029   \\
60359.21225(50)     & 0.145(16)    & 18878   & 0.00141    \\
\noalign{\smallskip}
\hline
\end{tabular}
\end{table}

\begin{figure}[t]
\centering
\includegraphics[width=84mm]{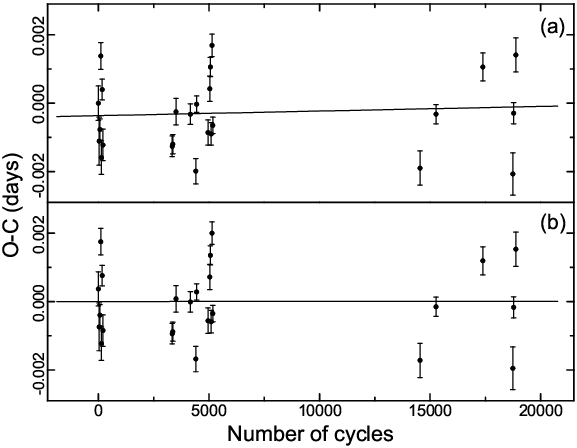}
\caption{(a) (O--C) diagram for the times of maxima calculated using the preliminary ephemeris shows a noticeable slope. (b) (O--C) diagram calculated using the corrected ephemeris does not show any slope.}
\label{fig6}
\end{figure}

It is widely accepted that, for eclipsing binary systems, the most precise way to determine the orbital period is by fitting an ephemeris to the times of eclipses. Because the 41-min oscillation is observed in each individual light curve, I can use the same approach to determine the oscillation period by using the times of maxima of the 41-min oscillation. Because the power spectra in Figures~\ref{fig2} and \ref{fig3} do not show high-frequency harmonics of the 41-min oscillation, this suggests that the pulse profile is similar to a sine wave. Therefore, to find the times of maxima and oscillation amplitudes, I used sine wave fits to each individual light curve. Beforehand, the light curves were averaged over 128-s time intervals. These times of maxima were referred to the middles of the light curves. The results are shown in Table~\ref{tab3}. The average semi-amplitude was found to be 0.161(6)~mag.

To obtain a preliminary ephemeris, I used the time of the maximum of the first light curve and the period measured using the power spectrum of the overall combined time series. Then, I calculated the observed minus calculated (O--C) values for the times of maxima in all light curves (Table~\ref{tab3}). The O--C diagram shown in Fig.~\ref{fig6}(a) has a notable slope. In addition, the O--C errors differ. Therefore, it is not appropriate to correct for this slope without considering weights. However, some points have deviations greater than $3\sigma$, indicating that the errors may be underestimated. Therefore, I increased all errors by a factor of 1.5 and used them as weights. As a result, I obtained a linear relationship: $O-C = -0.000\,37(15) + 0.000\,000\,013(18) {\it E}$. If I used the initial errors, the coefficients in this relationship would be the same and their errors would be a factor of 1.5 less. After correcting for the slope and the slight vertical offset seen in Fig.~\ref{fig6}(a), I obtained a final ephemeris with no slope or offset. The corrected O--C diagram is shown in Fig.~\ref{fig6}(b). The final ephemeris is: 

{\scriptsize
\begin{equation}
BJD_{\rm TDB}(\rm max) = 2459815.342\,53(15) + 0.028\,809\,630(18)*{\it E}.
\label{ephemeris1}
\end{equation} }

%ephem 2459815.34290(50) + 0.028809617(39)*
%A	-3.66124E-4=-0.00037    	1.50009E-4 = 0.00015
%B	1.3394E-8=0.000000013	1.82884E-8 = 0.000 000 018
%corr ephem 1.5sig     59 815.34253(15) + 0.028 809 630(18)

\begin{figure}[t]
\centering
\includegraphics[width=84mm]{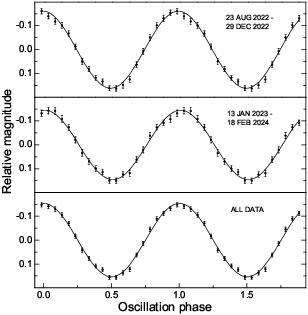}
\caption{Light curves of J0145 divided into two groups and folded with a period of 41.485867~min. The bottom frame shows all the light curves folded using this period. Solid curves show the fitted sine waves.}
\label{fig7}
\end{figure}

\begin{table}[t]
\caption[ ]{Periods obtained by different methods.}
\label{tab4}
\begin{tabular}{@{}l l c}
\hline
\noalign{\smallskip}
Method       &   Period (min)    & Dev.  \\
\noalign{\smallskip}
\hline
\noalign{\smallskip}
Average period from 25 power sp.      & 41.27(10)         &   2.2$\sigma$     \\
Averaged power spectrum                  & 41.41(13)         &   0.6$\sigma$    \\
Combined time series 2022 Aug.      & 41.488(12)       &   0.2$\sigma$    \\
Combined time series 2023 Jan.       & 41.486(9)         &   0.0$\sigma$    \\
Overall combined time series                  & 41.485849(56)  &   0.3$\sigma$    \\
Fitting the ephemeris           & 41.485867(26)  &   --        \\
\noalign{\smallskip}
\hline
\end{tabular}
\end{table}

The final oscillation period determined by the linear fit to the O--C values is 0.028\,809\,630(18)~d or 41.485\,867(26)~min. This period is close to the one obtained from the power spectrum of the overall combined time series, but with an error that is about two times less. It is believed that the validity time of an ephemeris is about four times shorter than the time during which the accumulated error of the period reaches one oscillation cycle (e.g., K. Mukai, {\small (\url{https://asd.gsfc.nasa.gov/Koji.Mukai/iphome/iphome.html})}. The accumulated error of the period determined using the linear fit to the (O--C) values reaches one oscillation cycle in 126~yr. Therefore, the rounded validity time of ephemeris~\ref{ephemeris1} is 30~yr. Table~\ref{tab4} shows the oscillation periods derived from different methods. Within measurement errors, these periods are consistent with each other.

%0.028 809 630(18) = 41.485867(0.000026) 56/26=2.15
%41.485867^2/0.000026=66 195 275 min=45 968.9 days=125.77 years /4= 31.4 years

Figure~\ref{fig7} shows the light curves of J0145 divided into two groups and folded with a period of 41.485867~min. The bottom frame shows all the light curves folded using this period. As seen, the pulse profile does not change throughout the observations. Furthermore, these light curves suggest that the pulse follows a perfect sinusoidal pattern. Indeed, only a few points deviate slightly more than 1$\sigma$ from the fitted sine waves, and these deviations are small. This sinusoidal profile confirms the absence of high-frequency harmonics in the power spectra. Using a sine wave fit to all the light curves, I found that the semi-amplitude of the pulse profile is 0.154(2)~mag. This agrees with the average semi-amplitudes obtained from individual power spectra (Table~\ref{tab2}) and from sine wave fits to individual light curves (Table~\ref{tab3}).

A characteristic feature of IPs is the presence of orbital sidebands. The power spectra in Figures~\ref{fig2} and \ref{fig3} do not show any additional prominent or repeated peaks that could be attributed to these sidebands. However, the power spectrum of the overall combined time series shown in Figure~\ref{fig4} has a complex structure around the main peak of the 41-min oscillation. Additional coherent oscillations with small amplitudes may be hidden within this structure. To find them, I used a well-known method involving subtracting the main oscillation from the data. An artificial overall combined time series was created, consisting of a sine wave with a period of 41.485867~min, and gaps corresponding to the gaps in observations. This time series, when folded with this period, showed the same amplitude and phase as the real overall combined time series. Then I subtracted the artificial overall time series from the real overall time series to obtain a pre-whitened combined time series that does not contain the 41-min oscillation.

\begin{figure}[t]
\centering
\includegraphics[width=84mm]{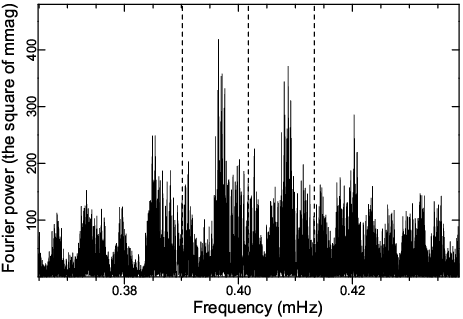}
\caption{Part of the power spectrum of the overall combined time series, from which the 41-min oscillation is subtracted. The dashed lines indicate the frequencies of the main peak of the 41-min oscillation and its one-day aliases.}
\label{fig8}
\end{figure}

The part of the power spectrum of the pre-whitened overall combined time series is shown in Figure~\ref{fig8}. As seen, the main peak of the 41-min oscillation and its one-day aliases disappeared. However, four clusters of peaks appeared, with the peaks distributed in frequency as one-day aliases. However, these peaks do not indicate the presence of an additional coherent oscillation, because the distribution of their amplitudes over frequency does not follow the window function shown in Figures~\ref{fig4} and \ref{fig5}. These peaks are the result of the superposition of noise and a weakly coherent oscillation, caused probably by the flickering of the star. This weakly coherent oscillation can be seen in the form of peaks located between the main peak and one-day aliases of the 41-min oscillation in the lower frame of Figure~\ref{fig3}. However, this weakly coherent oscillation is not present in the upper frame. Furthermore, the power spectra in Figure~\ref{fig3} do not show any repeated peaks at lower frequencies, which would indicate orbital periodicity. Thus, my extensive photometric observations of J0145 revealed only one coherent oscillation with a period of 41.485867(26)~min.

\section{Discussion}\label{sec4}

I conducted extensive photometric observations of J0145 over 25 nights, for a total of 139 hours. I clearly detected an oscillation with a period of 41.485867(26)~min. This period was similar to that previously reported by G.~Murawski for J0145, which had a value of 40.3229~min without specifying the measurement error (VSX database). It is difficult to determine the precision of Murawsky's period due to the lack of details in his work, but it is evident that I observed the same oscillation. My measurements had high precision due to the long duration of my observations, covering 18 months. This is the most significant finding in my work on J0145. In addition, the detected oscillation remained coherent throughout the entire observation time, spanning over 18 months - another important discovery.
%40.3229 min

Short-period X-ray oscillations that confidently classify J0145 as an IP have not yet been detected. According to \citet{kuulkers06}, the presence of additional orbital sidebands in addition to a stable short-period optical oscillation can be a strong argument for including J0145 in the list of IPs. Unfortunately, after long-term observations, I was unable to detect any additional oscillations in the power spectra of J0145 that could be associated with orbital sidebands. However, E.~Breedt obtained an optical spectrum of J0145, which showed distinct hydrogen and helium emission lines, typical of CVs with high accretion rates and in particular IPs (see {\small \url{https://vsx.aavso.org/vsx_docs/844443/4999/1RXS0145+5143.pdf}}). Based on this, it is suggested that J0145 is not an interacting binary white dwarf system with an ultra-short orbital period as such systems do not typically have hydrogen lines in their spectra (VSX database). Additionally, the observed 41-min oscillation cannot be attributed to a white dwarf pulsating non-radially. CVs with this type of pulsation typically exhibit clear signs of absorption lines from the white dwarf in their optical spectra, indicating a low relative contribution of flux from the accretion disc and the red dwarf \citep[e.g.,][]{nilsson06}. These absorption lines are not visible in the optical spectrum obtained by E.~Breedt. Therefore, the period of 41.485867(26)~min could probably be related to the spin period of a white dwarf in an IP. 

My detection of an oscillation with a period of 41.485867(26)~min provides independent confirmation of an oscillation previously discovered by G.~Murawski in J0145. In addition, I showed that this period is stable and the oscillation is highly coherent, which strengthens the probability that J0145 is an IP. However, it seems premature to consider J0145 as a confirmed IP. I studied the IP homepage by K. Mukai and the references therein to identify the main characteristics of ironclad and confirmed IPs. There are 71 such IPs in total, seven of which do not show short-period X-ray oscillations or additional optical oscillations associated with orbital sidebands. These seven IPs only show optical oscillations with short periods typical of IPs. These include: DQ~Her \citep[$P_{\rm spin}=71.065\,5828(3)$~s,][]{zhang95}, V455~And \citep[$P_{\rm spin}= 67.619\,685(40)$~s,][]{kozhevnikov15}, CXOGBS~J174954.5-294335 \citep[$P_{\rm spin}=503.32(3)$~s,][]{johnson17}, 1RXS~J230645.0+550816 \citep[$P_{\rm spin}=464.452(4)$~s,][]{halpern18}, V1460~Her \citep[$P_{\rm spin}=38.871\,256\,78(10)$~s,][]{pelisoli21}, PBC~J1841.1+0138 \citep[$P_{\rm spin}=311.805(4)$~s,][]{halpern22} and LAMOST~J024048.51+195226.9 \citep[$P_{\rm spin}=24.9328(38)$~s,][]{pelisoli22}. However, these seven IPs have known orbital periods, which is a significant characteristic of any binary star. Therefore, a short-period optical oscillation with a stable period should be sufficient to identify a CV as a confirmed IP, provided that the known orbital period rules out the possibility of the star being any other type besides a CV. 

Because my photometric observations did not detect any variability consistent with the potential orbital period of J0145, it can only be classified as a probable IP at this time. In order to classify J0145 as a confirmed IP, we need to determine its orbital period. This seems possible through future spectroscopic observations. J0145 is a relatively bright star of 16~mag, showing distinct emission lines in its spectrum and allowing measurements of radial velocity. 

My observations of J0145 also show another sign that it is a CV: the flickering that is typical of CVs. This flickering is clearly visible in the averaged power spectrum, where red noise extends up to frequencies of 20~mHz (Figure~\ref{fig2}). While atmospheric effects can contribute to red noise, the atmospheric red noise frequencies in my three-channel photometer are much lower, typically below a few mHz, as seen in Figure 2 in \citet{kozhevnikoviz00}. The detected flickering supports the hypothesis that J0145 is a CV and, therefore, an IP.

In most IPs, the period of the X-ray oscillation with the largest amplitude indicates that this period corresponds to the spin period of the white dwarf. This is because the X-ray sideband amplitudes are typically smaller. In contrast, optical spin oscillations in known IPs can have both larger and smaller amplitudes than the amplitudes of the optical orbital sidebands \citep{warner95}. Because short-period X-ray oscillations have not been detected in J0145, it is difficult to definitely conclude whether the period of 41.485867(26)~min is the white dwarf spin period or the orbital sideband period. However, pulse profiles of sideband oscillations often exhibit significant variability, and variations in the accretion disc structure are considered possible explanations for such variability \citep{woerd84}. Therefore, because the pulse profile of the 41-min oscillation showed no change over 18 months of my observations (see Figure~\ref{fig7}), it is likely that the period of 41.485867(26)~min is the white dwarf spin period rather than the orbital sideband period.

Whereas short-period oscillations are a definite characteristic of IPs, they are often difficult to detect and observe due to their small amplitudes. This is especially true if their semi-amplitudes are less than 0.01~mag. These oscillations can only be reliably detected by combining time series from several observation nights. In contrast, the average semi-amplitude of the 41-min oscillation in J0145 is very large, equal to 0.154(2)~mag. This semi-amplitude is one of the largest among known IPs. Indeed, the IP homepage by K. Mukai contains 58 confirmed and ironclad IPs with measured oscillation semi-amplitudes, and only four of these have semi-amplitudes greater than the semi-amplitude of the 41-min oscillation in J0145. These include: EX~Hya \citep[0.24~mag,][]{patterson94}, WX~Pyx \citep[0.20~mag,][]{donoghue96}, V1323~Her \citep[0.20~mag,][]{gansicke05} and FO~Aqr \citep[0.18~mag,][]{patterson94}. Due to the relatively low noise level, the oscillation period of J0145 can be measured with high precision. In addition, the 41-min oscillation has a sinusoidal pulse profile (Fig.~\ref{fig7}). This allows us to determine the oscillation phase with high precision using a sine wave fit. Therefore, J0145 is an excellent candidate for future studies of oscillation period changes.

Long-term tracking of the oscillation period in some IPs can be done through direct measurements of the period \citep[(e.g.,][]{patterson20}. However, if we assume that the white dwarf in J0145 has a d$P$/d$t$ of $10^{-11}$, which is typical for slowly spinning IPs \citep[see Table 1 in][]{warner96}, then the period change would be 0.000005~min per year, which is much less than the period error obtained during my 18-month observations (0.000026~min). Even after ten years, the expected change in the period will only be slightly larger than the error in determining the period. Therefore, despite the large oscillation amplitude, direct period measurements are not suitable for long-term tracking of the period in J0145. This type of measurement may be better suited to IPs with faster spin rates, as their periods can be determined with relatively high precision. For example, from 15-month observations of 1RXS~J230645.0+55081, with an oscillation period of 464.45600~s, I obtained a period error of 0.00010~s, which was three times less than the expected period change per year \citep{kozhevnikov22}. Therefore, direct measurements of the period over several years can provide statistically significant evidence of period change for 1RXS~J230645.0+55081.

%0.000005*10=0.00005
%0.000005*60=0.0003
%0.0001/60=0.0000016666666

Instead of directly measuring the period, the oscillation period of J0145 can be tracked over time by analyzing O--C values derived using ephemeris~\ref{ephemeris1}. To calculate the expected O--C values, assuming d$P$/d$t=10^{-11}$ for J0145, I used the formula provided by \citet{breger98}
\begin{equation}
{\rm (O - C)} = 0.5 \, \frac{1}{P} \, \frac{{\rm d}P}{{\rm d}t} \, t^2.
\label{breger}	
\end{equation}
Using formula~\ref{breger}, I calculated the expected O--C value to be 0.000023~d per year. 
However, the error in the O--C value obtained from a single observation night is significantly larger, ranging from 0.00024 to 0.00070~d (see Table~\ref{tab3}). Despite this, the expected O--C value according to formula~\ref{breger} is 0.0023~d over 10 years. This is 3--10 times larger than the error in the O--C value obtained from a single night, so we can detect changes in the oscillation period over this time if there are nonlinear changes in the O--C values. As shown in \citet{breger98}, this could occur if there are monotonic period changes. To make these non-linear changes detectable, it is necessary to conduct observations over a few nights per year. Then, using O--C values, long-term tracking of the oscillation period in J0149 could yield results after about 10 years of observations.

%Dp/dt=10^-11  ?P=dP/dt*?t=10^-11*365.25*86400 = 0.00032s = 0.000005 мин за год
%Dp/dt=10^-11  ?P=dP/dt*?t=10^-11*10*365.25*86400 = 0.0032s =  0.00005 мин за десять лет
% типичная ошибка Р 0.001-0.002с за сезон 0.0001/60=0.00000167min за два сезона
% типичная ошибка Р 0.00003-0.00006 мин за 18 мес. сравнима с изм. периода за 10 лет
%O-C=0.5*1/P*dP/dt*t^2=0.5/0.0053756481*10^-11*(365.25)^2=0.00012d =10c
%O-C=0.5*1/P*dP/dt*t^2=0.5/0.0053756481*10^-11*(10*365.25)^2=0.012d
%типичная ошибка О-С (0.05-0,07)*10^-3d=0.00005-0.00007d=4.3-6.0с 
%O-C=0.5*1/P*dP/dt*t^2=0.5/0.028809630*10^-11*(365.25)^2=0.000023d =10c за год
%O-C=0.5*1/P*dP/dt*t^2=0.5/0.028809630*10^-11*(10*365.25)^2=0.0023d = 200с за 10 лет
%типичная ошибка О-С за ночь (0.0003-0.0007)d  0.0023/0.0003=7.7     0.0023/0.0007=3.3
%типичная ошибка О-С за 10 ночей (0.0003-0.0007)/3.16=(0.0001-0.0002)d  0.0023/0.0001=23

\section{Conclusion}\label{sec5}

I conducted extensive photometric observations of J0149 over the course of three years from 2022 to 2024 for a total observation time of 139 hours spread over 18 months and 25 nights. These observations led to the following conclusions:

\begin{enumerate}
\item I clearly detected an oscillation with a period of about 40~min that was recently discovered by G.~Murawski in J0149 and reported in the VSX database. This oscillation was observed during both long and short observation nights.

\item The oscillation was coherent throughout all my observations over 18 months. This significantly increases the probability that J0149 is an intermediate polar

\item The extensive coverage of observations allowed me to precisely determine the oscillation period, which was found to be $41.485867\pm0.000026$~min.

\item The average oscillation semi-amplitude was very large, at 0.154(2) mag, and showed no systematic changes from year to year.

\item The oscillation pulse profile did not change throughout the observations and followed a perfect sinusoidal pattern.

\item I derived an oscillation ephemeris that is valid for 30~yr. This ephemeris can be used in future research to study changes in the oscillation period.

\end{enumerate}

%https://irsa.ipac.caltech.edu/Missions/ztf.html
%https://aladin.u-strasbg.fr/AladinLite

\section*{Acknowledgments}

The work of V. P.~Kozhevnikov was supported by the Ministry of science and higher education of the Russian Federation, agreement FEUZ-2023-0019. This work has made use of NASA's Astrophysics Data System Bibliographic Services, the VizeR catalogue access tool \citep{ochsenbein00} and the Aladin sky atlas developed at CDS, Strasbourg observatory, France \citep{boch14, bonnarel00}. This work has made use of data from the European Space Agency (ESA) mission $Gaia$ (\url{https://www.cosmos.esa.int/gaia}), processed by the $Gaia$ Data Processing and Analysis Consortium (DPAC, \url{https://www.cosmos.esa.int/web/gaia/dpac/consortium}). Funding for the DPAC has been provided by national institutions, in particular the institutions participating in the $Gaia$ Multilateral Agreement.

\subsection*{Author contributions}
I, Valerij P. Kozhevnikov, am the only author of the manuscript. All content of the manuscript was prepared by the author.
\subsection*{Financial disclosure}

None reported.

\subsection*{Conflict of interest}

The author declares that there is no potential conflict of interest.

%\nocite{*}% Show all bib entries - both cited and uncited; comment this %line to view only cited bib entries;
\bibliography{kozhevnikov_ms}%

%\section*{Author Biography}(if applicable)

%\begin{biography}{\includegraphics[width=60pt,height=70pt,draft]{empty}}%{\textbf{Author Name.} This is sample author biography text this is %sample author biography text this is sample author biography text this %is sample author biography text this is sample author biography text %this is sample author biography text this is sample author biography %text this is sample author biography text this is sample author %biography text .}
%\end{biography}

\end{document}